\documentclass[nofootinbib,aip,twocolumn,showpacs,amsmath,amssymb,floatfix,superscriptaddress]{revtex4}
\usepackage{graphicx}

\begin{document}

\title{Excitable wave patterns in temporal systems with two long delays}


\date{\today}

\author{Francesco~Marino}
\affiliation{CNR - Istituto Nazionale di Ottica, 
                largo E. Fermi 6, I-50125 Firenze, Italy}

\author{Giovanni~Giacomelli}
\affiliation{CNR - Istituto dei Sistemi Complessi, 
                via Madonna del Piano 10, I-50019 Sesto Fiorentino, Italy}

\begin{abstract}

\noindent Excitable waves arise in many spatially-extended systems of either biological, chemical, or physical nature due to the interplay between local reaction and diffusion processes. Here we demonstrate that similar phenomena are encoded in the time-dynamics of an excitable system with two, hierarchically long delays. The transition from 1D localized structures to curved wave-segments is experimentally observed in an excitable semiconductor laser with two feedback loops and reproduced by numerical simulations of a prototypical model. While closely related to those found in 2D excitable media, wave patterns in delayed systems exhibit unobserved features originating from causality-related constraints. An appropriate dynamical representation of the data uncovers these phenomena and permits to interpret them as the result of an effective 2D advection-reaction-diffusion process.

\end{abstract}

\maketitle

Excitable media are spatially-extended, nonequilibrium systems in which a localized nonlinear response can propagate throughout the space as an undamped solitary wave \cite{zykov0,meron_rev,murray}. 
The local dynamics is characterized by a state that is linearly stable, but susceptible to finite-amplitude perturbations.
The return to equilibrium entails a large excursion in the phase space corresponding to the emission of a spike of well-defined amplitude and duration. Such temporal behaviour in conjunction with a suitable transport process give rise to a wealth of spatiotemporal structures, including solitary pulses in one dimension, expanding targets \cite{chem,zhabo}, wave-segments \cite{zykov1,sakurai,dahlem2009,zykov2} and rotating spirals \cite{winfree} in two and scroll waves in three spatial dimensions, respectively. These patterns are among the most widespread examples of self-organization processes in active media. Excitable waves are found in chemical systems, the most familiar of which is the Belousov-Zhabotinsky reaction \cite{zhabo}, but also in the context of combustion theory \cite{palacios} and dendritic growth \cite{brower}. They play a fundamental role in the functional aspects of many biological systems, such as nerve and cardiac cells \cite{kenener,davidenko,sidorov}, and have been also observed in experiments on nematic liquid crystals \cite{frisch}, discharge plasmas \cite{dinklage} and semiconductor microcavities \cite{marino}. 

In all these cases the spatial coupling supporting wave propagation is provided by diffusion-like processes, and the wave dynamics is described in terms of reaction-diffusion equations \cite{murray,zykov0}.
Here, excitable wave patterns are demonstrated in a radically different scenario, i.e. in the dynamics of a purely temporal system with two, hierarchically long time-delays.

When the delay is much longer than any other characteristic time-scale, the dynamics of a delayed system can be mapped into an equivalent spatiotemporal representation (STR) (for a recent review see \cite{yanchuk2017}). In the single time-delay case, the STR is obtained by mapping a delay-time segment onto a pseudo-spatial cell and the index numbers of the subsequent delay cells into a pseudotime variable \cite{arecchi1992}. This approach had considerable success, as complex behavior in the time domain often result into simpler spatiotemporal patterns in the new representation
\cite{Giacomelli1996,Giacomelli2012,java2015,Giacomelli2013,Marino2014-2017,larger,garbin2015,Romeira2016,Faggian}, enforcing its physical validity.

The generalization of the STR to the case of multiple long-delays has been first tackled in Ref. \cite{Giacomelli2014-2015}: essentially new phenomena could take place owing to the higher number of pseudo-spatial dimensions involved. 
For instance, in Stuart-Landau models with two-delays spiral defects and defects turbulence are shown to occur \cite{Giacomelli2014-2015} and two-dimensional chimera states and dissipative solitons have been recently observed \cite{Brunner2018}. 

In this Letter, we investigate the effects of two, hierarchically long delayed feedback loops on an excitable semiconductor laser. The temporal dynamics is shown to encode the transition from 1D localized structures to expanding wave segments. These travelling waves emerge as the strenght of the two feedbacks becomes comparable, and represent the 2D generalization of propagating pulses in this class of systems. While possessing many properties of excitable waves, the segments present negative curvature and remains confined to a well defined propagation cone. By means of a proper representation introduced in \cite{DR} here generalized and applied to experimental data, we construct an advection-reaction-diffusion model of our system where the observed wave dynamics is reproduced in the limit of strong advection.

\begin{figure}[t!]
\includegraphics[width=1.0\linewidth]{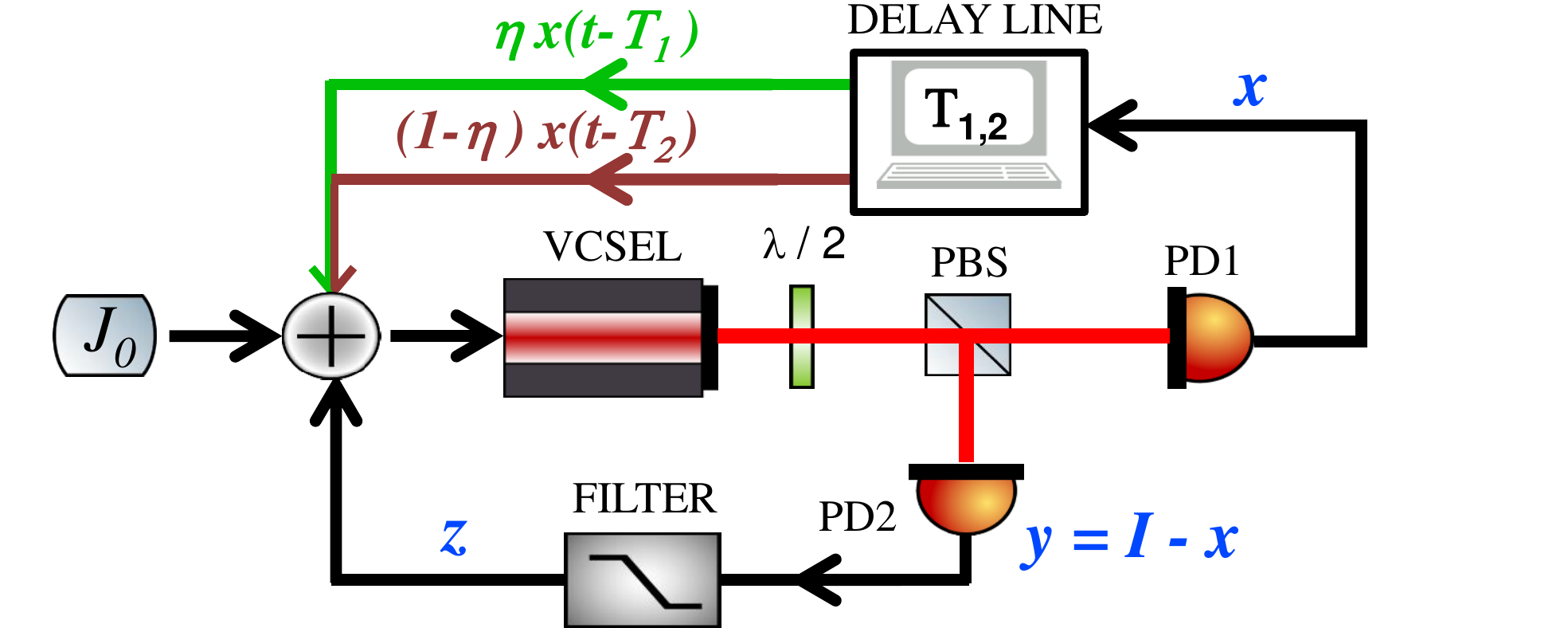}
\caption{Experimental setup. VCSEL: vertical-cavity surface-emitting laser; $\lambda/2$: half-wave plate; PBS: polarizing beamsplitter; PD1,2: photodetector(s). The delayed feedback is realized sampling the electric signal from the detector PD1 with a A/D-D/A board hosted by a PC driven by a real-time Linux OS.}
\label{fig1}
\end{figure}

The experimental setup is shown in Fig.\ref{fig1}. A Vertical Cavity Surface Emitting Laser (VCSEL) is operated in a regime of polarization bistability. The two linear polarizations are separated by means of a half-wave plate and a polarizing beamsplitter and then their intensities are monitored by photodetectors (1 MHz-bandwidth). Transitions between the two states and an hysteresis cycle are observed as the pump current $J_0$ is slowly varied. The signal of one polarization is low-pass filtered (filter cut-off around $1$~kHz) and summed to the DC pump current. As a result, whenever a polarization switch occurs, the current will start a slow evolution (with a characteristic time-scale $T_0$ in the $ms$ range determined by the low-pass filter) leading to the emission of an excitable spike \cite{Marinochaos}. The signal of the other polarization is sampled and retarded with delays $T_1 \gg T_0$ and $T_2 \gg T_1$ by a re-configurable A/D-D/A board with a bandwidth of 50 kHz. The two delayed output are then recombined with weights $\eta$ and $1-\eta$ and fed back into the VCSEL through the pump current with a gain $g$. 

\begin{figure}[t!]
\includegraphics[width=1.0\linewidth]{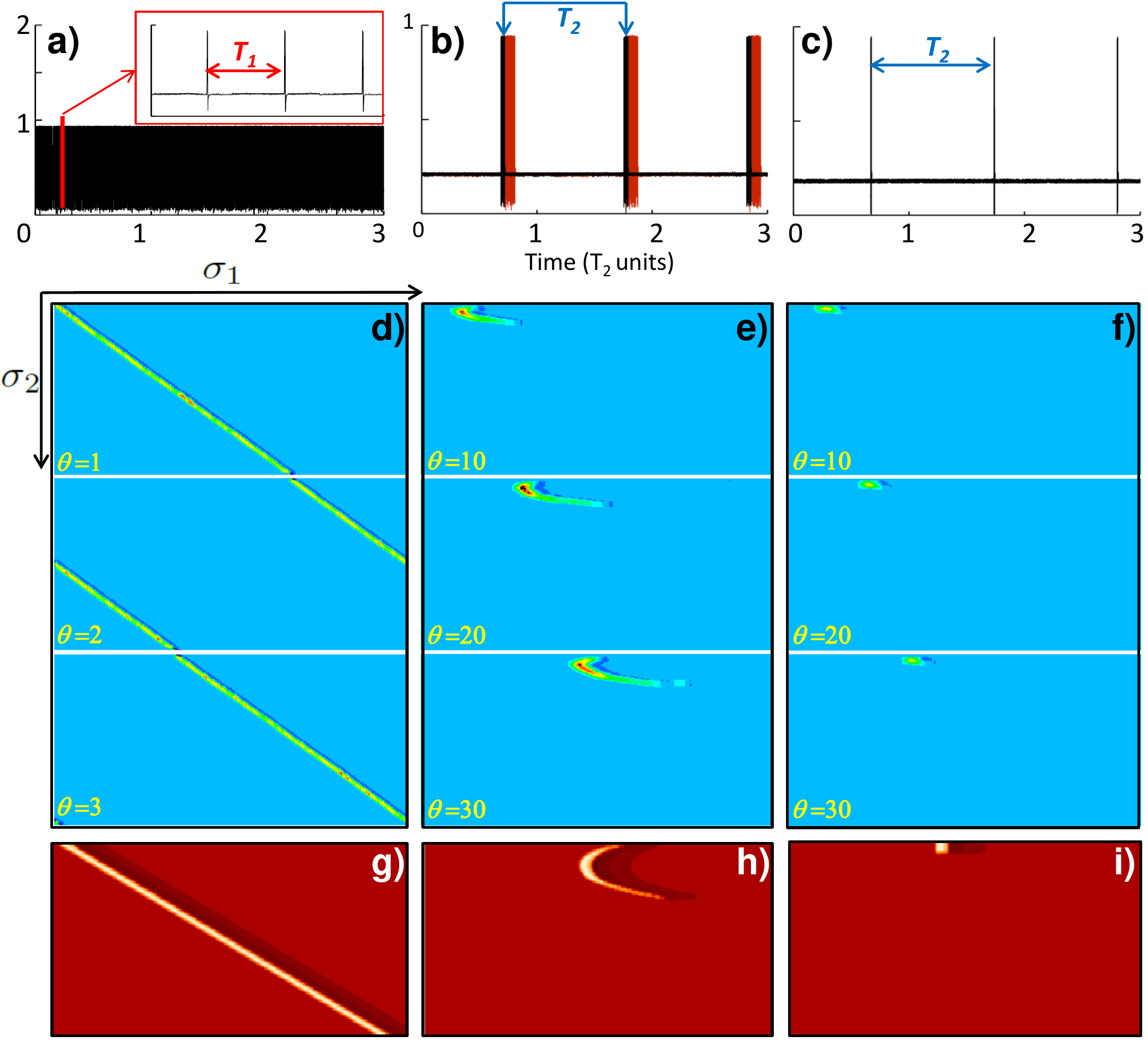}
\caption{Experimental time-series and snapshots of the STR of the laser intensity for different values of the asymmetry: a),d) $\eta=$ 0.9; b),e) $\eta=$ 0.4, c),f) $\eta=$ 0.1. b) The two sets of bursts plotted as dark and red (light) are snapshots of duration $3~T_2$ separated by $10~T_2$. The pump current is fixed at $7.2$~mA. The delays are $T_1$=$91$~ms and $T_2$= $20.9$~s. Snapshots of the STR as obtained by numerical integration of Eqs. (\ref{model2}) for g) $\eta=$ 0.9, h) $\eta=$ 0.4, i) $\eta=$ 0.1. Other parameters: $J=$~-0.65, $g=$ 0.2$, \alpha=$1.5, $\varepsilon$=$0.05$. See videos in the supplemental material for this and the following figures.}
\label{fig2}
\end{figure}

The dynamics of the system can be described by a simple prototypical model \cite{Marinochaos}, here generalized to the case of two delayed feedbacks 
\begin{eqnarray}
\dot{x} &=& F(x) + J_0 + \alpha z + g_1 x_{T_1} + g_2 x_{T_2} + \zeta \nonumber \\
\dot{z} &=& -\varepsilon~(z - y),
\label{model}
\end{eqnarray}
where $x_{T_i} = x(t- T_i)$, $i=1,2$ are the delay terms, $J_0$ is the DC pump current and $\{x(t),y(t)\}$ are proportional to the polarizations signals. As such, the relation $x(t)+y(t)=I$ approximately holds, where $I$ is proportional to the photo-current relative to the total intensity of the laser. $z(t)$ is a low-pass filtered function (with a cut-off frequency $\varepsilon$) of $y$ that is added to the bias current with the coupling coefficient $\alpha$. The nonlinear function $F(x)= x - x^3$ is a phenomenological potential governing the mode-hopping between polarization modes \cite{Giacomelli2012,VanExter98}, $g_1 = g \eta$ and $g_2 = g (1 -\eta)$ are the gain coefficients of the feedback loops and $\zeta$ is a $\delta$-correlated, white Gaussian noise. 
By introducing the new parameter $J = J_0+\alpha I$ and the variable $w=J_0 + \alpha z$, Eqs. (\ref{model}) become
\begin{eqnarray}
\dot{x} &=& F(x) + w + + g_1 x_{T_1} + g_2 x_{T_2} + \zeta \nonumber\\
\dot{w} &=& -\varepsilon~(w - J + \alpha x).
\label{model2}
\end{eqnarray}
The model (\ref {model2}) with $g_{1,2}$ = 0 has the form of the well known FitzHugh-Nagumo (FHN) equations \cite{fhn}, one of the paradigmatic models displaying excitability.

In Fig. \ref{fig2}a-c we show three time-series of the optical intensity for different values of the asymmetry $\eta$ and the same gain. For $\eta \approx 1$, the feedback term with delay $T_1$ is dominant. As a result, when the system is initialized in a in-homogeneous initial condition, we observe a periodic sequence of spikes with period close to $T_1$ (see Fig. \ref{fig2}a). Conversely, at low $\eta$ values, the period of the pulse train approaches $T_2$ (see Fig. \ref{fig2}c). When both delayed terms are significant, an intermediate regime takes place where spikes separated from each other by nearly $T_1$ are emitted in bursts with a period $T_2$ (Fig. \ref{fig2}b).
Unlike the cases shown in Figs. \ref{fig2}a,c where the waveforms are periodic, here the number of spikes forming each burst increases with time (red trace in Fig. \ref{fig2}b) thus reducing the duty cycle of the sequence. This typically occurs over time-scales of several $T_2$.

The corresponding spatiotemporal plots are obtained through the transformation
\begin{equation}
t = \sigma_1  +  \sigma_2 T_1 + \theta T_2~,
\label{STR}
\end{equation} 
which generalizes the STR for systems with two, hierarchically long-delays, $T_0\!\ll \!T_1 \!\ll \!T_2$  \cite{Giacomelli2014-2015}. The time-series are cut into consecutive segments of length $T_2$, each labeled by the integer $\theta$, and further divided into smaller intervals of lenght $T_1$ and index number $\sigma_2 \! \in \! \left[0, T_2/T_1 \right)$. In this representation, $\sigma_1 \! \in \! \left[0, T_1 \right)$ and $\sigma_2$ act as pseudo-spatial variables, while $\theta$ plays the role of the pseudo-time coordinate \cite{Giacomelli2014-2015}.

In Fig.\ref{fig2}d we show the STR corresponding to the regime shown in Fig. \ref{fig2}a. Since $g_2 x_{T_2} \ll g_1 x_{T_1}$, the time variable can be written as $t \approx \sigma_1  +  \sigma_2 T_1$, with $\{\sigma_1,\sigma_2\}$ acting as pseudo-space and -time respectively. We thus expect the dynamics to be essentially confined in one spatial dimension. What is in fact shown in Fig. \ref{fig2}d is nothing but a different representation of the patterns observed in excitable systems with a single time-delay \cite{garbin2015,Romeira2016,Marinochaos}, where 1D localized structures propagate through the $(\sigma_1,\sigma_2)$ pseudo-spacetime and $\theta$ only arbitrarily labels snapshots over intervals $T_2$. An analogous situation occurs at small $\eta$ values, where the localized structure now lies in the orthogonal $\sigma_1-\theta$ plane. The spot propagating along $\sigma_1$ shown in Fig. \ref{fig2}f, thus corresponds to a stroboscopic mapping of the dynamics with a period $T_2$. 

As the two feedback terms become comparable, 1D localized structures evolve into 2D wave patterns: the STR indeed shows a curved wave segment (Fig. \ref{fig2}e). The local excitation spreads across the pseudo-space indicating the existence of an effective 2D spatial coupling mechanism.

In Fig. \ref{fig2}g,h,i we report the numerical simulations of Eq. (\ref{model2}) for parameters corresponding to the experimental conditions. The results agree with the observations, suggesting that the above findings are indeed generic features of delayed excitable systems.

\begin{figure}[t!]
\includegraphics[width=1.0\linewidth]{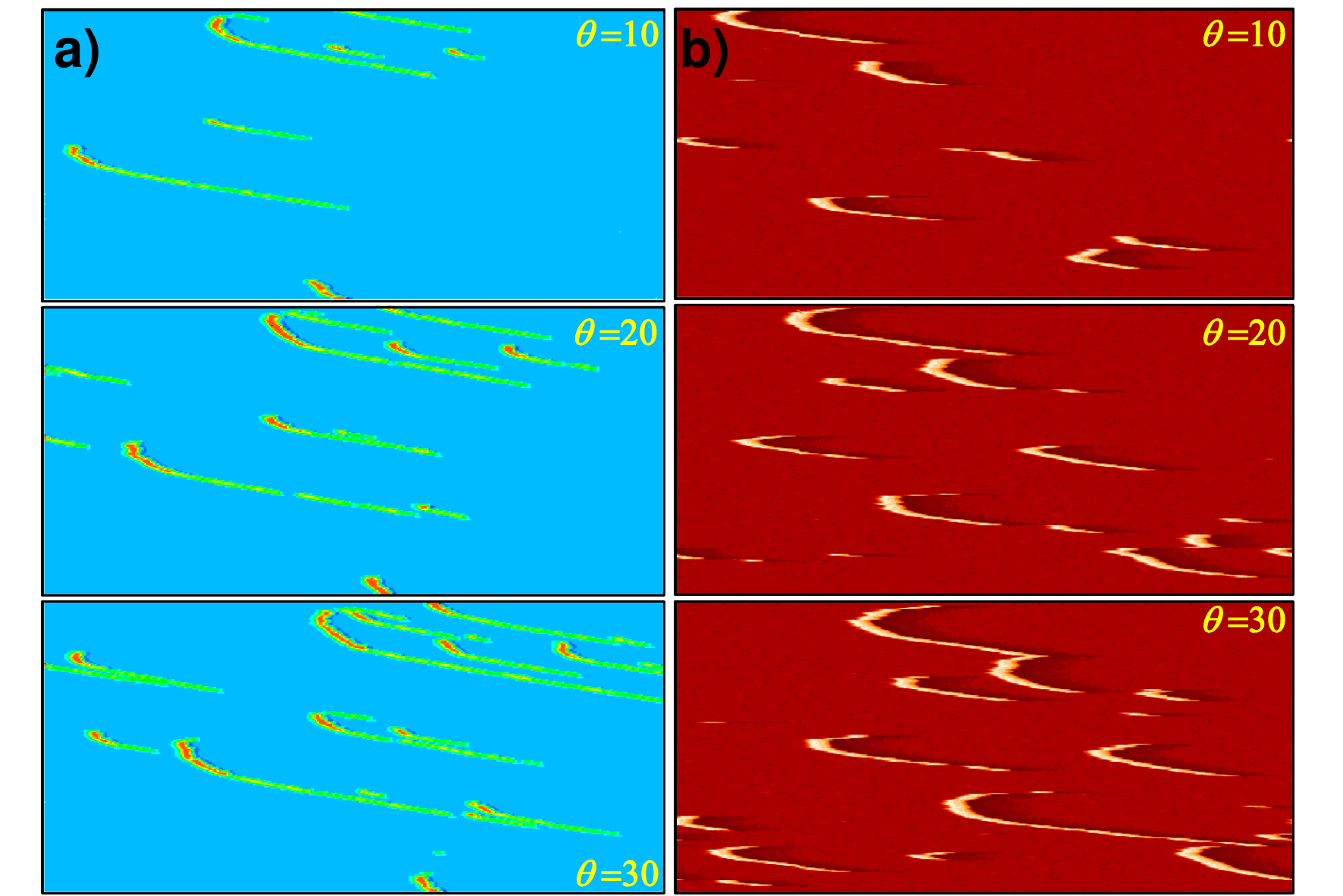}
\caption{Snapshots in the STR showing the noise-induced generation, propagation and interaction of excitable wave segments. a) Experiment: the pump current is $7.35$~mA. b) Simulation of model (\ref{model2}) with a noise amplitude of $\tau=10^{-2}$. Other parameters as in Fig. \ref{fig2}.}
\label{fig3}
\end{figure} 

For higher pump values, corresponding to a lower excitability threshold, noise fluctuations sporadically trigger the excitations (see Fig. \ref{fig3}a-b). The resulting wave segments propagate without decrement with a speed that is insensitive to initial conditions and annihilate after collision (due to refractoriness).

These features are immediately reminiscent of wave propagation processes in reaction-diffusion media \cite{zykov0,chem}. Nevertheless, here the patterns present remarkable differences. In 2D homogeneous media the propagation takes place symmetrically around the origin of the excitation, leading to the emission of a target wave. In fact, we observe excitations in the form of negatively-curved (concave) wave segments. While wave-fronts and even wave segments with negative curvature have been observed hitherto in spatially-extended media, they typically arise in peculiar situations, e.g. when convex waves collide \cite{foerster,wussling} or in media with a negative effective diffusion \cite{maree}.

\begin{figure}[t!]
\includegraphics[width=1.0\linewidth]{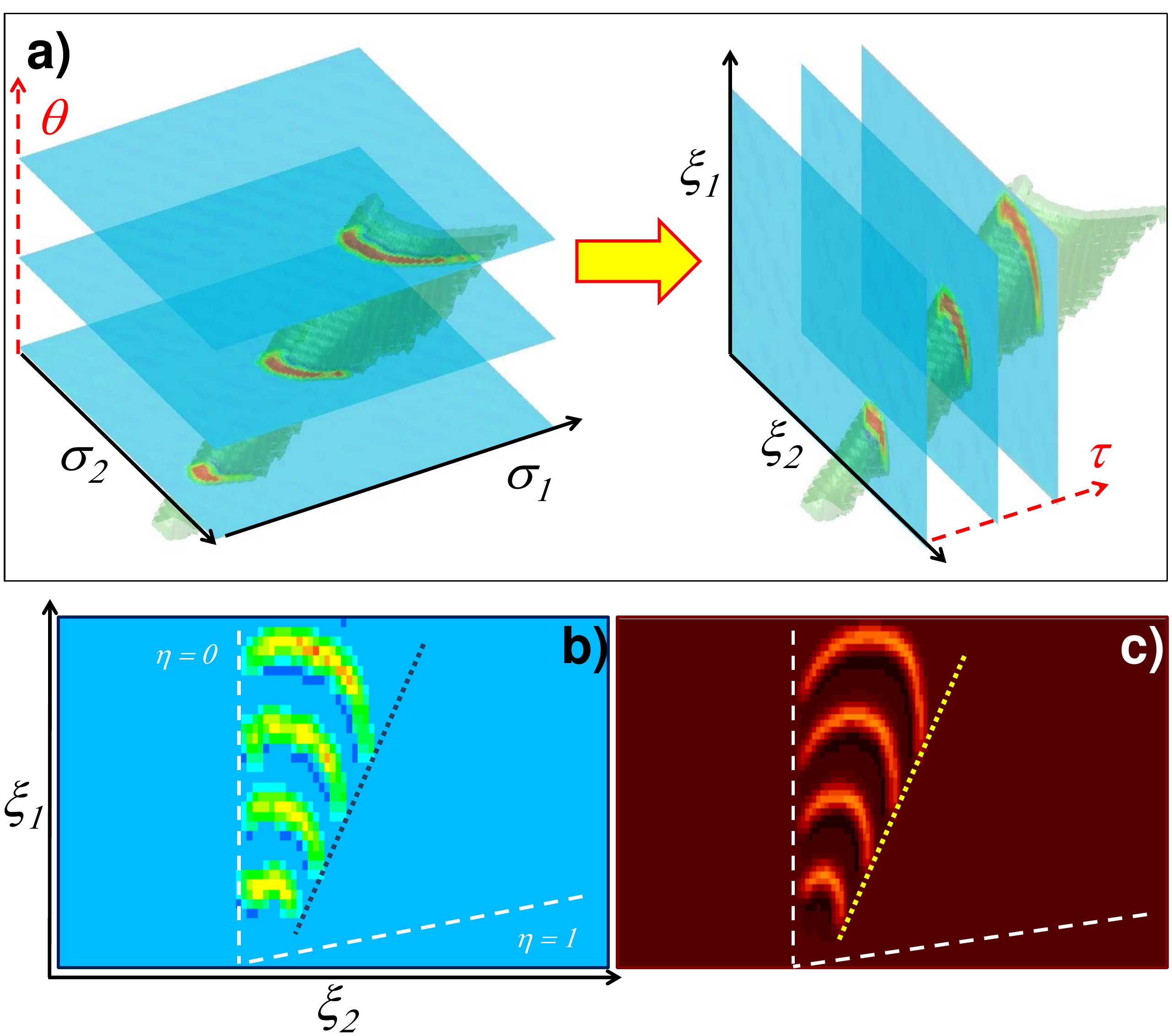}
\caption{(a) Spatiotemporal representations. Experimental isosurface extracted from data showing the propagation of a single wave segment as in Fig. \ref{fig2}. Horizontal cuts, perpendicular to the $\theta$-direction correspond to snapshots in the STR while vertical image planes correspond to snapshots in the DR. b) Snapshots of experimental (every 0.39~ms) and c) numerical (every 25 $\tau$-steps) wave-segments propagation in the new representation. Parameters as in Fig. \ref{fig2}.}
\label{fig4}
\end{figure}

Such an anomalous behavior can be elucidated by means of the recently introduced dynamical representation (DR) \cite{DR}, here generalized to the case of two delays. The DR employs an alternative definition of equivalent space and time variables with respect to the STR, considering $\{\sigma_2,\theta\}$ as spatial and $\sigma_{1}$ as temporal variables respectively. We denote the corresponding spatial and temporal variables as $\{\xi_1, \xi_2\}$ and $\tau$. The two representations are visualized in Fig. \ref{fig3}a, where we plot as an example the isosurface extracted from measurements of a single wave-segment.
The orientation of the image planes cutting the data identifies the direction of evolution and thus the two different representations. Cuts perpendicular to the $\theta$ direction correspond to the standard STR, in which wave segments possess a negative curvature (left panel). The patterns evolve over the $\theta$ direction on a domain with almost periodic boundaries on $\sigma_{1,2}$ which, in analogy to spatially-extended systems, lead to the commonly adopted identification of $\sigma$ and $\theta$ as pseudo-space and -time respectively. The DR is shown in Fig \ref{fig3}a (right panel), where the vertical image planes now identify $\tau$ as the new temporal axis.
Although the boundary conditions of the STR appear as the most natural, it has been shown that the bulk dynamics (i.e. far from boundaries) is more properly obtained in the new representation \cite{DR}. Wave segments indeed become convex as commonly observed in reaction-diffusion models. This is clearly evidenced in Fig. \ref{fig4}b) where four $\tau$-snapshots of the dynamics in the $\xi_1-\xi_2$ pseudo-space are plotted. The figure also reveals that the wave-segment remains confined to a specific propagation cone. The dashed oblique and vertical lines corresponds to the trajectories followed by the 1D localized structures shown in Fig 2d) and 2f) respectively, and both relate to causality. The vertical line marks a causal boundary of the pseudo-spacetime, as by construction no world line could ever cross it independently on the parameters or system under consideration. The line for $\eta = 1$ is instead related to the intrinsic drift present in long-delayed systems, which again is due to causality \cite{yanchuk2017} as evidenced by the maximal comoving Lyapunov exponents \cite{Giacomelli1996,Giacomelli98}. Such drift depends on the feedback gain $g$. Finally, the dotted line indicate the trajectory of a 1D localized structure obtained setting $g_2=0$ and $\eta=0.4$, which is the asymmetry parameter corresponding to the 2D patterns shown in the figure.

Remarkably, the DR provides an explicit rule for generating the equivalent spatio-temporal dynamics. Using Eq.(\ref{STR}) re-written in the DR and defining  
$\{u(\xi_1, \xi_2, \tau),v(\xi_1, \xi_2, \tau)\} = \{x(t),w(t)\}$, Eqs.(\ref{model2}) become
\begin{eqnarray}
\partial_\tau u &=& F(u)+ v + g_1 u(\xi_1, \xi_2 -1, \theta)
+ g_2 u(\xi_1 -1, \xi_2, \tau) \nonumber\\
\partial_\tau v &=& -\varepsilon~(v - J + \alpha u)~,
\label{model_sp}
\end{eqnarray}
where the delayed terms turn into {\it non-local} asymmetric, spatial couplings which break the $\{\xi_1,\xi_2\}$ symmetries. A description in terms of partial differential equations can be thus obtained by {\it formally} expanding the non-local terms as
\begin{eqnarray}
u(\xi_1-1,\xi_2,\tau) &\approx& u(\xi_1,\xi_2,\tau) -\partial_{\xi_1}u(\xi_1,\xi_2,\tau) \nonumber\\ 
&&+{1\over 2}\partial^2_{\xi_1 \xi_1}u(\xi_1,\xi_2,\tau)-...~,
\label{expansion}
\end{eqnarray}
and similarly for $u(\xi_1,\xi_2-1,\tau)$. As discussed in \cite{DR} the validity of the expansion (\ref{expansion}) relies on the {\it a-posteriori} examination of the dynamics generated by Eq.(\ref{model2}), since the scale of the evolution cannot be generally determined in advance. 
Expanding the delayed terms up to second-order we obtain
\begin{eqnarray}
\partial_\tau u &=& (1 + g)u - u^3 +v - ({\bf v_d} \cdot {\bf \nabla}) u +  {\bf \nabla} \cdot ( {\bf{\mathcal{D}}} {\bf\nabla} u) \nonumber\\
\partial_\tau v &=& -\varepsilon~(v - J + \alpha u)~,
\label{model_st}
\end{eqnarray}
where ${\bf \nabla}$=($\partial_{\xi_1}$, $\partial_{\xi_2}$), ${\bf v_d}$=($g_1$,$g_2$) and $\mathcal{D}$=$diag(g_1/2,g_2/2)$ is a 2$\times$2 diagonal matrix. Eqs. (\ref{model_st}) is a spatially extended FHN model with advection and anisotropic diffusion. Except for anisotropy, the normal velocity of a wave-front depends linearly on its local curvature and the flow velocity and on the square root of the diffusion coefficient via the eikonal equation \cite{foerster,zykov2,foerster2}. Here, both the flow vector ${\bf v_d}$ and the diffusion tensor $\mathcal{D}$ depend on the gain coefficients $g_{1,2}$. For high values of $g$ we thus expect the wave propagation to be mainly determined by the advection term and front curvature. 

\begin{figure}[t!]
\includegraphics[width=1.0\linewidth]{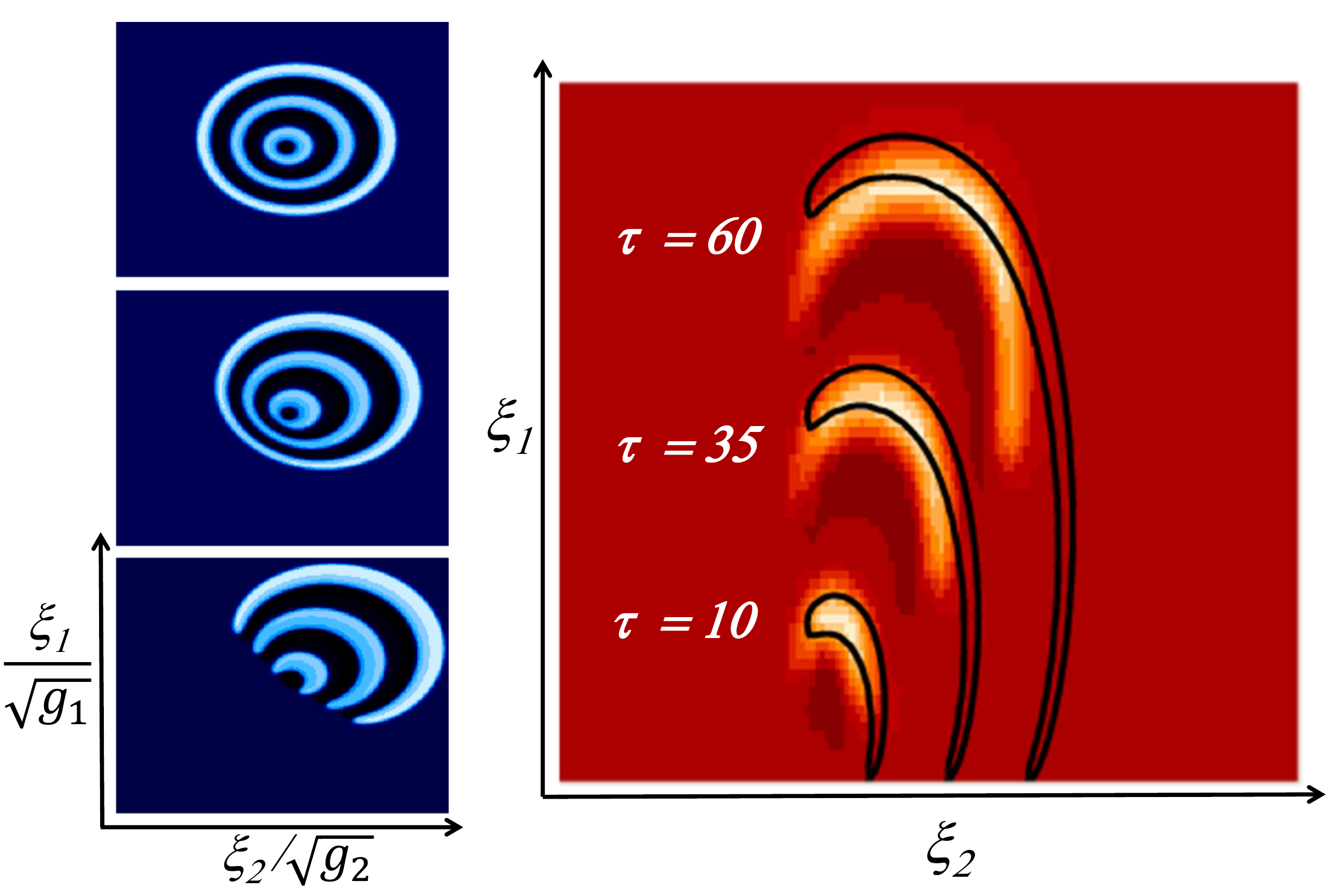}
\caption{Left panels: snapshots of the excitable wave dynamics at times $t=$ 25, 55, 85  obtained from Eqs. (\ref{model_st}) for $\eta$=0.5. From top to bottom: $g=$ 0.01, $g=$ 0.08, $g=$ 0.2. Right panel: snapshots of wave-segments for $g=$ 0.2 and $\eta$=0.4 from  model (\ref{model2})
(see also Fig. \ref{fig4}) and from Eqs. (\ref{model_st}) (black contour curves at half maximum).}
\label{fig5}
\end{figure}

In Fig. \ref{fig5} (left) we report the excitable wave dynamics as obtained from Eqs. \ref{model_st} for three
values of the parameter $g$ and isotropic diffusion ($\eta=0.5$).
At sufficiently low values of $g$, we observe the propagation of a target-wave as in 2D reaction-diffusion systems with no advection. A substantially different behavior is found when $g$ is increased.
The wave is initially deformed, with a wavefront velocity that is larger in the direction of the flow and then, beyond a critical value of the gain the front propagating upstream is suppressed leading to wave segment traveling in the opposite direction. Similar phenomena have been previously reported in reaction-diffusion models with a shear flow \cite{biktashev,neufeld}. 

In Fig. \ref{fig5} (right) we compare the patterns obtained from the delay model (\ref{model2})
and the spatially extended system (\ref{model_st}). While some discrepancies are found at their extrema,
the waves show a good qualitative agreement not only for velocity, but also for duration and curvature.
On the other hand, targets and their transition to wave-segments have never been observed either in the delayed model (\ref{model2})  or experimentally for any value of the system parameters. This is due to the fact that the delay non-locality induces a symmetry-breaking which confines the propagation to specific spatial regions. The  second order expansion of the nonlocal terms in model (\ref{model_st}) does not breaks instead the spatial symmetry, as the advection could be removed in the co-moving reference frame. 

In conclusion, we have shown both experimentally and theoretically the occurrence of traveling wave segments in an excitable system with two, hierarchically long delays. The waves emerge as the two delay terms become comparable, i.e. in the presence of a two-dimensional pseudo-space and thus represent the 2D analogue of one-dimensional propagating pulses. These phenomena are disclosed by means of a proper dynamical representation, here generalized to higher dimensions and tested with experimental data. Remarkably, the reconstruction naturally gives rise to a spatially-extended model of the dynamics thus establishing a link with standard descriptions. Our results open the way to further studies on reaction-diffusion waves in delay setups, analogous e.g. to in-homogeneous media, involving higher pseudo-spatial dimensions or curved surfaces \cite{dahlem2014}. We also expect that the possibility of finely tuning the effective spatial dimensionality of the system could unveil new pattern-formation and self-organization processes.

\end{document}